\documentclass[9pt,conference]{IEEEtran}

\usepackage[preprint]{waspaa25}

\usepackage{bm} %
\usepackage{algorithm}
\usepackage{algorithmic}

\definecolor{zncolor}{rgb}{0.7,0.1,0.4}
\definecolor{jpcolor}{RGB}{255, 0, 255}
\definecolor{tbkcolor}{RGB}{0,100,50}
\definecolor{jmcolor}{rgb}{0.7,0.3,0.7}

\newcommand{\ourmethod}{{ARC} post-training}
\newcommand{\ourmethodfull}{\textbf{A}dversarial \textbf{R}elativistic-\textbf{C}ontrastive post-training}

\title{Fast Text-to-Audio Generation with Adversarial Post-Training}

\name{Zachary Novack$^{1,2*\dagger}$, Zach Evans$^{2*}$, Zack Zukowski$^{2}$, Josiah Taylor$^{2}$, CJ Carr$^{2}$, Julian Parker$^{2}$}{Adnan Al-Sinan$^{3}$, Gian Marco Iodice$^{3}$, Julian McAuley$^{1}$, Taylor Berg-Kirkpatrick$^{1}$, Jordi Pons$^{2}$}
\address{$^{1}$UC -- San Diego, $^{2}$Stability AI, $^{3}$Arm, $^*$Shared authorship, $^\dagger$Work done while an intern at Stability AI}

\begin{document}

\maketitle

\begin{abstract}
Text-to-audio systems, while increasingly performant, are slow at inference time, thus making their latency unpractical for many creative applications. We present Adversarial Relativistic-Contrastive (ARC) post-training, the first adversarial acceleration algorithm for diffusion/flow models not based on distillation. While past adversarial post-training methods have struggled to compare against their expensive distillation counterparts, \ourmethod{} is a simple procedure that (1) extends a recent \emph{relativistic} adversarial formulation to diffusion/flow post-training and (2) combines it with a novel \emph{contrastive} discriminator objective to encourage better prompt adherence. We pair \ourmethod{} with a number optimizations to Stable Audio Open and build a model capable of generating $\approx$12s of 44.1kHz stereo audio in $\approx${75ms} on an H100, and $\approx$7s on a mobile edge-device, the fastest text-to-audio model to our knowledge.
\end{abstract}

\section{Introduction}
\label{sec:introduction}

Despite the recent progress building generative text-to-audio models, %
such systems require seconds to minutes \emph{per generation}~\cite{copet2023simple,stableaudio,Evans2024LongformMG,evans2024open}, which limits their usability for most creative use cases.
Our goal is to accelerate gaussian flow-based models, \textit{i.e.},~diffusion models~\cite{song2020denoising} or the more recent rectified flows~\cite{esser2024scaling}. %
Since gaussian flow models can be costly at inference time due to their iterative (step-based) sampling~\cite{ho2020denoising}, considerable effort has gone into accelerating them~\cite{yin2023one,yin2024improved,Sauer2024FastHI,Sauer2023AdversarialDD}. %
Most of these works rely on (step) distillation, where the teacher model provides direct supervision to train a distilled few-step generator. 
As such, the few-step generator learns to map multiple inference steps into a single step, or a small number of steps, by distilling the teacher’s trajectories.
However, most distillation approaches come with practical drawbacks: like
\emph{online} methods~\cite{Ren2024HyperSDTS, Wang2024PhasedCM, song2023consistency, lu2024simplifying, chen2025sana, Kim2023ConsistencyTM, Novack2025Presto, xu2025one, yin2024improved}, which are costly to train as they require 2-3 full models %
held in memory at the same time, or \emph{offline} methods~\cite{liu2022flow, salimans2022progressive, kang2024distilling}, which require significant resources to generate and store trajectory-output pairs offline to later train on.
{Furthermore}, most distillation setups~\cite{Ren2024HyperSDTS, Wang2024PhasedCM, song2023consistency, lu2024simplifying, Kim2023ConsistencyTM, Novack2025Presto} distill the teacher with Classifier-Free Guidance (CFG), which is a double-edged sword, as the generator inherits both the high quality/prompt adherence as well as the low diversity and over-saturation artifacts of CFG \cite{chung2024cfgpp}.

To avoid such drawbacks, some studied acceleration through \emph{post-training} (without distillation)~\cite{xu2024ufogen, lin2025diffusion}. These works are {primarily} adversarial, as opposed to distillation methods that use adversarial auxiliary losses~\cite{yin2024improved,Novack2025Presto,Kim2023ConsistencyTM,Sauer2024FastHI,Sauer2023AdversarialDD}, and use {real} data rather than teacher-generated samples, thus freeing the costly requirement of using trajectory-output pairs. 
The motivation 
here
is that 
the adversarial loss encourages \textit{realism}, %
making each estimate better than the standard gaussian flow estimate.
Such improved estimates enable post-trained models to use fewer (1-8) sampling steps~\cite{xu2024ufogen, lin2025diffusion}.
Recent image works, like UFOGen~\cite{xu2024ufogen} and APT~\cite{lin2025diffusion}, explored adversarial post-training, yet UFOGen reported limited gains~\cite{xu2024ufogen} and APT required a distillation-based initialization~\cite{lin2025diffusion}.
How to improve adversarial post-training and apply it to audio remains an open question.

\begin{figure}
    \centering
    \includegraphics[width=0.35\textwidth,clip]{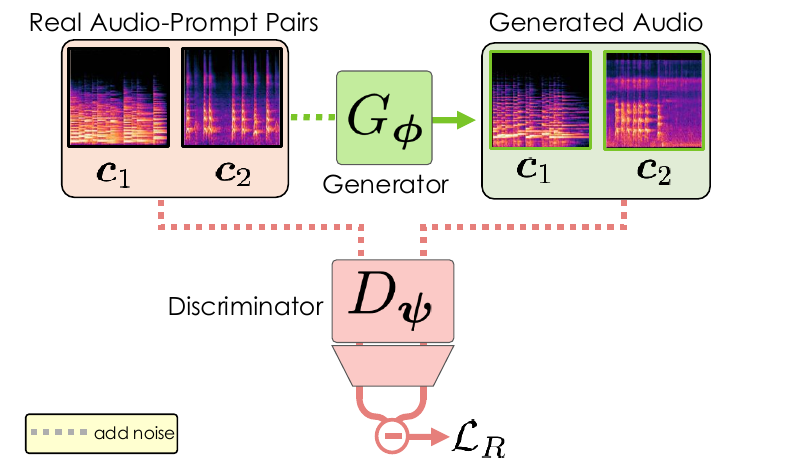}
    \vspace{-0.9em}
    \caption{\textbf{ARC's adversarial relativistic loss}. Pairs of generated and real samples (with the same text prompts) are passed into the discriminator (with additive noise), where the generator and discriminator are trained to minimize and maximize (respectively) the difference between fake and real outputs.}
    \label{fig:main_figr}
    \vspace{-1.7em}
\end{figure}
\begin{figure}
    \centering
    \includegraphics[width=0.35\textwidth,clip]{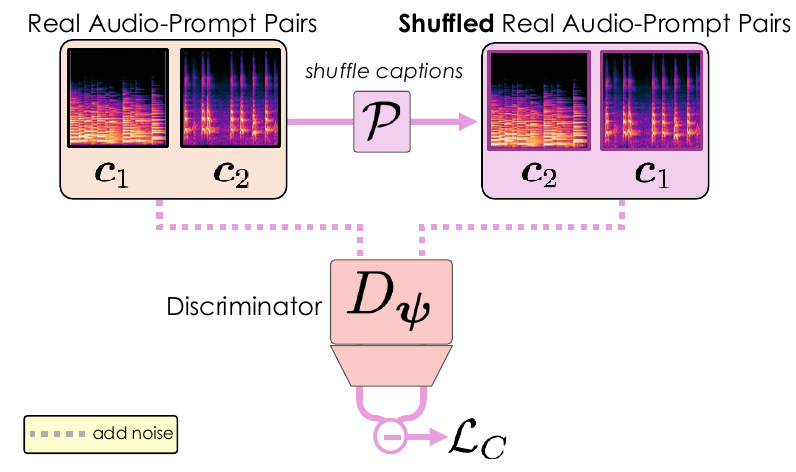}
    \vspace{-0.9em}
    \caption{\textbf{ARC's contrastive loss}. The discriminator is also trained to maximize the difference %
    between audios with correct and incorrect (shuffled) {prompts}.}
    \label{fig:main_figc}
    \vspace{-2em}
\end{figure}

Within the audio-domain, only a few works exist on accelerating gaussian flow models. Most rely on consistency models~\cite{bai2023accelerating, Saito2024SoundCTMUS, nistal2024improving, Novack2024DITTO2DD} and have no adversarial component, despite the presence of adversarial losses in their image-domain counterparts~\cite{Kim2023ConsistencyTM}. 
The only work to use any adversarial component is Presto~\cite{Novack2025Presto}, although it is mainly based on distillation~\cite{yin2024improved}. Finally, most existing accelerated text-to-audio models focus on band-limited~\cite{bai2023accelerating}, mono\cite{Saito2024SoundCTMUS,Novack2024DITTO2DD, Novack2025Presto} audio.

We show that rectified flows can be accelerated with adversarial post-training, introducing the first fully-adversarial text-to-audio acceleration framework that requires neither distillation nor CFG.
We propose \textbf{A}dversarial \textbf{R}elativistic-\textbf{C}ontrastive post-training (\textbf{ARC}) that (1)~extends a recently proposed \emph{relativistic} adversarial formulation~\cite{huang2024gan,jolicoeur2018relativistic} {to text-conditional audio generation} and (2)~combines this with a novel \emph{contrastive} loss that encourages the discriminator to focus on prompt adherence.
To our knowledge, we are the first to attempt such a post-training recipe on audio generation, the first to use adversarial relativistic losses for gaussian flow acceleration across modalities, and the first to use a contrastive discriminator formulation.
We pair ARC with an array of architecture updates to Stable Audio Open (SAO) \cite{evans2024open} to achieve accelerated performance, %
generating $\approx$12s of 44.1kHz {stereo} audio in $\approx${75ms} on a H100 GPU, being 100x faster than the original SAO model.  Across experiments, we find that ARC is competitive with existing state-of-the-art acceleration methods, and notably preserves generative diversity far beyond previous methods.
We further optimize it for on-device CPU-based inference, achieving a text-to-audio model that can run locally on edge-devices (\textit{i.e.},~smartphones) with an inference time of $\approx$7s. To our knowledge, we are the first to operate at this inference speed, enabling efficient on-device execution, and facilitating the applicability of text-to-audio in creative domains. We make \href{https://github.com/Stability-AI/stable-audio-tools/blob/main/stable_audio_tools/training/arc.py}{our code} public, and a \href{https://arc-text2audio.github.io/web/}{demo website} is available.

\section{METHOD}
\label{sec:method}

\subsection{Rectified flows pre-training}
\label{sec:rf}
Given a prompt $\bm{c}$ and a (stereo audio) representation $\mathbf{x}_0$, the goal of (text-to-audio) rectified flows \cite{liu2022flow, esser2024scaling} is to learn a model that transfers between the data distribution $p_0$ and some prior distribution $p_1$ (\textit{e.g.}, isotropic gaussian noise) given $\bm{c}$, thus allowing to generate samples from $p_0$ by sampling from $p_1$ given $\bm{c}$ according to the learned model. The forward corruption process (from data to noise) is given by:
\begin{equation}\label{eq:fm}
    \mathbf{x}_t = (1-t)\mathbf{x}_0 + t\bm\epsilon, \quad \bm\epsilon \sim \mathcal{N}(0, \bm{I}),
\end{equation}
where $t$ is the scalar {time} parameter ranging from 0 (data) to 1 (noise) {that is $p_\text{gen}(t)$ distributed}. For brevity, we refer to the noising process in Eq. \ref{eq:fm} as sampling from the forward probability distribution $q(\mathbf{x}_t \mid \mathbf{x}_0)$. To reverse this process and generate samples (from noise to data), we can solve the ordinary differential equation (ODE):
\begin{equation}\label{eq:ode}
    \mathrm{d}\mathbf{x}_t = -{v}_{{\theta}}(\mathbf{x}_t, t, \bm{c}) \mathrm{d}t,
\end{equation}
where ${v}_{{\theta}}({\mathbf{x}_t}, t, \bm{c})$ is a model trained to predict the {velocity} of the flow $\mathbf{v} = \bm\epsilon - \mathbf{x}_0$. This model can thus be trained by adding noise to real samples according to Eq.~\ref{eq:fm} and predicting the velocity:
\begin{equation}\label{eq:rfloss}
    \arg\min_{\bm\theta}\mathbb{E}_{\mathbf{x}_0, \bm{c} \sim \mathcal{X}, \bm{\epsilon} \sim \mathcal{N}(0, \bm{I}), t \sim p_\text{gen}(t)} [\|\mathbf{v} - {v}_{{\theta}}(\mathbf{x}_t, t, \bm{c})\|_2^2],
\end{equation}

\subsection{\ourmethodfull{} (ARC)}
\label{sec:rcapt}

Gaussian flow models require tens to hundreds \cite{esser2024scaling} of steps to produce high-fidelity outputs, as the objective in Eq.~\ref{eq:rfloss} ($\ell_2$-based)
is minimized by
estimating the instantaneous velocity of the flow at any noise level, which is equivalent to fitting the
noise-conditional \textit{mean} 
of the denoised outputs.
This results in sampling schedules that require a number of small sampling steps to maintain stable sampling trajectories~\cite{frans2024one}.
Thus, the goal of adversarial post-training is to turn our pre-trained flow model $v_{\bm\theta}$ into a few-step generator $G_{\bm\phi}$ by
supplanting the $\ell_2$-based conditional \textit{mean} loss with an adversarial loss, defined by some discriminator $D_{\bm\psi}$. %
Such a discriminator evaluates the \emph{realism} of denoised samples, providing distribution-level feedback that goes beyond the broad conditional \textit{mean} loss. 
As such, if the denoised output at any given step is sufficiently \textit{real} and higher-quality, fewer steps are required.
A key advantage of adversarial post-training over (the costly) distillation methods is that it avoids the generation and storage of trajectory-output pairs for \textit{offline} methods and storing 2-3 full generative models in memory for \textit{online} methods.
It also sidesteps reliance on the potentially poor performance of the pre-trained (teacher) model.
While adversarial post-training is a broad class of algorithms, our method is based on combining a \textit{relativistic} adversarial loss and a \textit{contrastive} loss that are jointly optimized: %
\vspace{-0.15em}
\begin{equation}\label{eq:both}
     \min_{\bm\phi}\max_{\bm\psi}\mathcal{L}_\text{ARC} (\bm\phi, \bm\psi) = \mathcal{L}_\text{R} (\bm\phi, \bm\psi) + \lambda \cdot \mathcal{L}_\text{C} (\bm\psi),
\end{equation}
where $\mathcal{L}_\text{R}$ and $ \mathcal{L}_\text{C}$ are defined in Sec. \ref{sec:rapt} and \ref{sec:capt}, respectively.

\subsection{Adversarial relativistic loss: the AR in ARC}
\label{sec:rapt}

To start,
both the few-step generator $G_\phi(\mathbf{x}_t, t, \bm{c})$ and the discriminator $D_{\bm\psi}(\mathbf{x}_t, t, \bm{c})$ are initialized from ${v}_{{\theta}}(\mathbf{x}_t, t, \bm{c})$
that is $t$ and $\bm{c}$ conditioned and already can process {noisy} data $\mathbf{x}_t$.
This initialization improves training stability.
Under this setup, adversarial relativistic post-training proceeds as in Fig. \ref{fig:main_figr}: given $\mathbf{x}_0$ (clean data) and $\bm{c}$ (text prompt), we corrupt $\mathbf{x}_0$ with noise following Eq.~\ref{eq:fm} to obtain $\mathbf{x}_t$ (noised data) for some $t$ sampled from {$p_{\text{gen}}(t)$.
This is then passed into the generator to produce $\mathbf{\hat{x}}_0 = G_{\bm\phi}(\mathbf{x}_t, t, \bm{c})$\footnote{In practice, for training stability we do not convert the model parameterization to predict the clean output directly, and instead keep the base velocity prediction and parameterize the generator as $G_{\bm\phi}(\mathbf{x}_t, t, \bm{c}) = \mathbf{x}_t - t v_{\bm\phi}(\mathbf{x}_t, t, \bm{c})$.}.
Given the generated $\mathbf{\hat{x}}_0$ and the real $\mathbf{x}_0$, we again add noise following Eq.~\ref{eq:fm}\footnote{The $\epsilon$ noise is independently sampled for each element of the real/generated pairs, but the noise level $s$ (or $t$ in Eq.~\ref{eq:fm}) is fixed per pair.} given some timestep $s$ that is sampled from {$p_{\text{disc}}(s)$}. Then, the noisy samples $\mathbf{x}_s$ and $\mathbf{\hat{x}}_s$ become inputs to the discriminator $D_{\bm\psi}$.
Although past work uses the same distribution for $p_{\text{gen}}(t)$ and $p_{\text{disc}}(s)$, we disentangle these two for added flexibility and performance~\cite{Novack2025Presto}.
We train both the discriminator and generator with this \emph{relativistic} adversarial loss~\cite{huang2024gan}:
\begin{align}\label{eq:rgan}
     &\mathcal{L}_\text{R} (\bm\phi, \bm\psi) =\underset{\substack{\mathbf{x}_0, \bm{c} \sim \mathcal{X}\\ t \sim p_{\text{gen}}(t),\\s \sim p_{\text{disc}}(s)}}{\mathbb{E}} \big[{f}\big(\bm{\Delta}_{\text{gen}}(\mathbf{x}_0, t, s, \bm{c}) - \bm{\Delta}_{\text{real}}(\mathbf{x}_0, s, \bm{c})\big)\big] \\
    &\bm{\Delta}_{\text{gen}}(\mathbf{x}_0, t, s, \bm{c}) = D_{\bm\psi}(q(\mathbf{\hat{x}}_s \mid G_{\bm\phi}(q(\mathbf{x}_t \mid \mathbf{x}_0) , t, \bm{c})), s, \bm{c})\\
        &\bm{\Delta}_{\text{real}}(\mathbf{x}_0, s, \bm{c}) = D_{\bm\psi}(q(\mathbf{x}_s \mid \mathbf{x}_0), s, \bm{c}), 
\end{align}
where $f(x) = - \log (1 + e^{-x})$, $\bm{\Delta}_{\text{real}}$ are the discriminator logits on noisy versions of real samples, and $\bm{\Delta}_{\text{gen}}$ are the logits on noisy versions of generated samples, themselves generated from noisy versions of real samples. 
This {relativistic} adversarial loss is notably different from the standard GAN objective. 
The $G_{\bm\phi}$ of (standard) GANs aims to minimize the detection probability by $D_{\bm\psi}$,
while the $D_{\bm\psi}$ attempts to {maximize} the detection probability for generated samples and minimize detection probability for real samples---where each min/max happens independently.
Instead, $\mathcal{L}_\text{R}$ is calculated on \emph{pairs} of real/generated data \cite{huang2024gan,jolicoeur2018relativistic}, such that $G_{\bm\phi}$ minimizes its detection \textit{relative} to its paired real sample in  $D_{\bm\psi}$ space (with same prompt), %
and $D_{\bm\psi}$ {maximizes} detecting the generated sample \textit{relative} to its real pair. Thus, $G_{\bm\phi}$ wants every generated sample to be ``more real than its paired real sample", while  $D_{\bm\psi}$ wants every real sample to be ``more real than its paired generated sample". {Critically, while Huang et al.~\cite{huang2024gan} do not specify how samples may be paired, our pairs are always \emph{highly} related due to our text-conditional task, where pairs of real/generated samples share the same prompt $\bm c$, thus providing a stronger gradient signal than relying on random pairings.}

While we find {relativistic} losses more effective than standard GAN losses (Table \ref{tab:main}),
we also find that adversarial losses \emph{alone} are worse at prompt adherence than distillation.
Without direct distillation from a teacher, with strong prompt adherence through CFG~\cite{chung2024cfgpp}, 
it is hard for the $G_{\bm\phi}$ to improve prompt adherence with a realism-focused $D_{\bm\psi}$.

\subsection{Contrastive loss: the C in ARC}
\label{sec:capt}

 The above {relativistic} objective contributes in improving output quality and sampling efficiency, see Sec. \ref{sec:rcapt}, but it alone does not fix the poor text-following caused by the adversarial loss, see Sec. \ref{sec:rapt}. %
{Instead of using CFG to increase consistency, for simplicity, we again rely on a new relativistic loss using the same discriminator. Specifically,} we find that prompt adherence can be improved by training the discriminator also as an audio-text {contrastive} model~\cite{wu2023large}. %
Since our discriminator is natively text-conditional due to its pre-trained text-to-audio backbone, we train it with a similar {relativistic} loss %
that maximizes
the difference between real samples with 
incorrect and correct {prompts} (Fig. \ref{fig:main_figc}):
\begin{align}
     \mathcal{L}_\text{C}(\bm\psi) = \underset{\substack{\mathbf{x}_0, \bm{c} \sim \mathcal{X},\\s \sim p_{\text{disc}}(s)}}{\mathbb{E}} \Big[f\big(\bm{\Delta}_{\text{real}}(\mathbf{x}_0, s, \mathcal{P}[\bm{c}]) -
\bm{\Delta}_{\text{real}}(\mathbf{x}_0, s, \bm{c}) \big)\Big],
\end{align}
where the discriminator $D_{\bm\psi}$ 
maximizes the difference
between the 
incorrect and correct 
prompts in $D_{\bm\psi}$ logit space, and $\mathcal{P}[\cdot]$ denotes a random intra-batch permutation of the text prompts $\bm{c}$.
In words, by maximizing this objective, $D_{\bm\psi}$ attempts to make its outputs higher for
{incorrect audio-text pairs} \emph{relative} to the pairs with correct prompts.
This loss can be viewed as a %
contrastive loss~\cite{gutmann2010noise}, where the discriminator is mapping %
correct audio-text pairs closer than mismatched pairs.
Note that this is only a loss for the {discriminator}, as this encourages it to understand the alignment between prompts and noisy inputs and focus on semantic features, and prevents the model from focusing on easier (\textit{e.g.}, high-frequency \cite{Novack2025Presto}) 
features.
Thus, with a semantically-aware $D_{\bm\psi}$, the {relativistic} adversarial loss is tailored towards improving prompt adherence. %
Interestingly, $\mathcal{L}_C$ eliminates the need for CFG, which helps with prompt adherence but is documented to worsen diversity and over-saturate outputs~\cite{chung2024cfgpp}.

\subsection{Ping-pong sampling}
\label{sec:sampling}

With {ARC} post-training, our model is finetuned to directly estimate clean outputs $\mathbf{x}_0$ from different noise levels $\tau$, instead of predicting the instantaneous velocity like rectified flow models do. To adjust for that, we employ ping-pong sampling \cite{song2023consistency} instead of the traditional ODE solvers that the rectified flow models use (\textit{e.g.}, Euler). Ping-pong sampling alternates between denoising and re-noising to iteratively refine samples. Given some starting noisy sample $\mathbf{x}_{\tau_i}$, we denoise it using our few step generator $\hat{\mathbf{x}}_0 = G_{\bm\phi}(\mathbf{x}_{\tau_i} , {\tau_i}, \bm{c})$. Then, we {re-noise} the sample to some lower noise level $x_{\tau_{i-1}} = (1-\tau_{i-1})\hat{\mathbf{x}}_0 + \tau_{i-1}\epsilon$, where $\tau_{i-1} < \tau_{i}$ and $\epsilon \sim \mathcal{N}(0, \bm{I})$. This process repeats $N$ times, progressively improving the sample towards the final clean data $\hat{\mathbf{x}}_0$.

{Note that here we do not use CFG \cite{chung2024cfgpp}. Provided that CFG is memory intensive, because two forward passes (conditional and unconditional runs) are required \cite{chung2024cfgpp}, this decision is critical for our on-device experiments, as our method uses only half the (V)RAM.}

\subsection{Acceleration as reward modeling: a post-training perspective}

We can connect {ARC} to language models {preference} post-training (alignment)~\cite{ziegler2019fine, rafailov2023direct, shao2024deepseekmath}, wherein one trains a reward model on human preferences {(\textit{i.e.},~winning-losing pairs $\bm{y}_w, \bm{y}_l$ given the \textit{same} prompt $\bm{c}$)} to fit a preference model $p(\bm{y}_w \succeq \bm{y}_l \mid \bm{c})$ \cite{bradley1952rank}, and then post-trains (aligns) a language model to maximize these rewards. 
Specifically, because our relativistic objective is on \emph{pairs} of real/generated data given the \emph{same} text prompts $\bm{c}$, then with $f(x) = - \log (1 + e^{-x})$~\cite{huang2024gan} $\mathcal{L}_\text{R}$ is {equivalent} to maximizing the likelihood of $p_{\bm\psi}(\mathbf{x}_s \succeq \mathbf{\hat{x}}_s \mid \bm{c})$ as in language models. This means that $D_{\bm\psi}$ is trained implicitly as a reward model, where the reward is defined as \textit{realism} against an online-created dataset of generated samples, and our generator is trained to produce samples that maximize the relative reward $p_{\bm\psi}(\mathbf{\hat{x}}_s  \succeq\mathbf{x}_s \mid \bm{c})$. %

\begin{table*}[ht!]
    \centering
        \caption{\textbf{Qualitative and quantitative results} ($\downarrow$ the lower the better or $\uparrow$ the higher the better). For MOS scores we also present 95\% confidence intervals.}
    \label{tab:main}
    \resizebox{0.99\width}{!}{%
    \begin{tabular}{l|c||ccc|ccc|ccc|cc}
    \toprule
&  &  FD$_{\text{openl3}}$ & KL$_{\text{passt}}$&  CLAP &  CCDS & $R_{\text{passt}}$ & $C_{\text{passt}}$ & Diversity & Quality  &  Prompt ($\uparrow$)&  RTF  & VRAM \\
        
        Method& Steps & ($\downarrow$)&  ($\downarrow$)&  ($\uparrow$)& ($\uparrow$)& ($\uparrow$)& ($\uparrow$)& ($\uparrow$) &  ($\uparrow$)&   adherence &   ($\uparrow$)& (GB, $\downarrow$)\\
        \midrule

SAO \cite{evans2024open}  & 100 & {78.24}& {2.14}& {0.29}& {0.35}& {0.26}& {0.41} & {4.0}{\tiny $\pm$0.2}& \textbf{4.0}{\tiny $\pm$0.3}& \textbf{4.3}{\tiny $\pm$0.3}& 3.56&5.51\\
SAO ($\downarrow$ steps)  & 50 & 82.17& 2.22& 0.28& 0.34& 0.26& {0.42}& - &- &- & 7.01& 5.51\\
SAO ($\downarrow$$\downarrow$ steps)  & 8 & 143.21& 4.24& {\color{blue}{0.05}} & {\color{blue}{0.60}}& 0.17& 0.09&  - &- &- & 37.58& 5.51\\

\midrule
Pre-trained RF & 50& 88.13& \textbf{2.04}& \textbf{0.30}& 0.34& \textbf{0.34}& \textbf{0.42} & 3.1{\tiny $\pm$0.2}& 3.7{\tiny $\pm$0.3}& {4.2}{\tiny $\pm$0.2}& 18.82& 4.20\\
Pre-trained RF ($\downarrow$ steps)& 8& 91.97& 2.59& 0.23& 0.38& 0.25& 0.30& - &- &- & 100.74& 4.20\\
 $+${ARC} (ours) & 8 & \textbf{84.43}& 2.24& 0.27& \textbf{0.41}& 0.28& 0.37 & \textbf{4.4}{\tiny $\pm$0.3}& 3.5{\tiny $\pm$0.4}& 3.8{\tiny $\pm$0.5}& 156.42&4.06\\
 $+$Presto \cite{Novack2025Presto}& 8 & 93.05& {2.11}& 0.27& {\color{red}0.26}& 0.27& 0.36 & {\color{red} 2.7{\tiny $\pm$0.4}}& \textbf{4.0}{\tiny $\pm$0.3}& {4.2}{\tiny $\pm$0.3}& 156.42 & 4.06 \\
 $+$$\mathcal{L}_R$ \hspace{7.1mm}(w/o $\mathcal{L}_C$)& 8 & 90.92& 2.62& {\color{blue}0.20} & {\color{blue}{0.57}}& 0.29& 0.28 & - &- &-& 156.42&4.06\\
 $+$$\mathcal{L}_{\text{LS}}$$+$$\mathcal{L}_C$ (w/o $\mathcal{L}_R$)& 8 & 98.13& 2.39& 0.26& 0.36& 0.17& 0.34 &  - &- &-& 156.42&4.06\\ 
  \midrule
RF $+$ ARC (one-step) & 1 & 100.17& 2.45& 0.24& 0.33& 0.11&0.27& - & - &-&  440.30& 4.06\\  
RF $+$ ARC (few-step) & 4 & 90.45& 2.21& 0.27& 0.40& 0.26&0.40& - & - &-&  247.67& 4.06\\

     \bottomrule
    \end{tabular}
    }

\end{table*}

\section{Experiments}
\label{sec:experiments}

\subsection{Models}
\label{sec:model}

\hspace{2.5mm} \textit{Generative models} ($v_{\bm\theta}$, $G_{\bm\phi}$). Our latent generative models synthesize variable-length (up to 11.89s, with a timing control \cite{evans2024open}) stereo audio at 44.1kHz from text. It consists of the pre-trained 156M parameter {autoencoder} from SAO \cite{evans2024open} that compresses waveforms into a 64-channel 21.5Hz latent space, a 109M parameter T5 text embedder \cite{raffel2020exploring}, and a Diffusion Transformer (DiT) that operates in the latent space. Note that the DiT is first pre-trained as a rectified flow. 
Our model is a variant of SAO \cite{evans2024open} with improved efficiency: we reduce DiT's dimension from 1536 to 1024, the number of layers from 24 to 16, we add QK-LayerNorm \cite{henry2020query}, and remove the ``seconds start" embedding. These reduce the DiT from 1.06B to 0.34B parameters. %
During inference, we compile the base DiT with \texttt{torch.compile}.

\textit{Discriminator} ($D_{\bm\psi}$). Our discriminator is initialized using the weights of the pre-trained rectified flow \cite{Novack2025Presto, yin2024improved}. %
Specifically, our discriminator is initialized using the input embedding layers and 75\% of the DiT blocks of the pre-trained rectified flow. Then, these intermediate diffusion features are passed into a (randomly initialized) lightweight discriminator head, comprised of 4x blocks of 1D convolutions, interleaved with GroupNorm~\cite{wu2018group} and SiLU activations~\cite{elfwing2018sigmoid}. In total, the discriminator has $\approx274$M parameters.%

\subsection{Training and sampling details}\label{sec:tsdetails}

Our training data follows SAO \cite{evans2024open} but excludes the long-form FMA music, as our focus is on sound effects and loops. This results in 6,330h (472,618 audios) of Freesound samples with CC0, CC-BY, or CC-Sampling+ licenses. Our rectified flow model was trained for 670k iterations. For each acceleration algorithm, we finetune the model 100k iterations with a batch size of 256 across 8 H100 GPUs. We use a learning rate of $5$$\times$$10^{-7}$ for both the generator and discriminator, with AdamW, and set $\lambda$$=$$1$. %
$p_{\text{gen}}(t)$ is a uniform distribution in log-SNR space from -6 to 2~\cite{Novack2025Presto}, to align post-training with inference, while $p_{\text{disc}}(s)$ is the shifted logit normal distribution~\cite{esser2024scaling} which puts more weight on mid-to-high SNR regions~\cite{Novack2025Presto}. During \ourmethod{}, we alternate between updating $G_{\bm\phi}$ ($\mathcal{L}_R$) and updating $D_{\bm\psi}$ ($\mathcal{L}_R$+$\mathcal{L}_C$).%

\subsection{Objective Evalaution}

We use FD\textsubscript{openl3}~\cite{cramer2019look}, KL\textsubscript{passt}~\cite{koutini2021efficient}, and CLAP score~\cite{wu2023large} metrics, which are established metrics %
that broadly assess audio quality, semantic alignment, and prompt adherence. We also assess {diversity}, as recent text-to-image acceleration works noted a distinct diversity reduction in distilled models \cite{kang2024distilling, gandikota2025distilling}. Following previous audio works, we also report recall and coverage metrics~\cite{Novack2025Presto, nistal2024diff}, $R$\textsubscript{passt} and $C$\textsubscript{passt}, which measure the overall distributional diversity in PASST \cite{koutini2021efficient} space. Yet, there exists no metric for assessing \emph{conditional} diversity, \textit{i.e.},~the diversity of generations with the same prompt. %
To solve this, we %
propose the {CLAP Conditional Diversity Score} (or CCDS), which is calculated as the average CLAP cosine distance between pairs of generations within some batch with the same prompt $\bm{c}$, averaged across all batches.
Intuitively, if the distance between generations is low CCDS indicates low diversity. Accordingly, higher CCDS results indicate more diverse outputs.
Finally, to measure speed, we report Real-Time Factor (RTF),~which is the amount of audio generated divided by the latency, and VRAM peak usage measured on an H100. 
We evaluate our models with AudioCaps \cite{kim2019audiocaps} test set, that contains 979 audio segments, each with several captions (881 audios were available and it includes 4,875 captions).
We generate an audio per caption, resulting in 4,875 generations for FD\textsubscript{openl3}~\cite{cramer2019look}, KL\textsubscript{passt}~\cite{koutini2021efficient}, CLAP score~\cite{wu2023large}, $R$\textsubscript{passt}, and $C$\textsubscript{passt}. For CCDS we generate 24 audios per caption from a subset of 203 randomly selected AudioCaps prompts.

\subsection{Subjective Evaluation}

We also run a listening test with webMUSHRA \cite{schoeffler2018webmushra}. Participants were asked to rate diversity, audio quality, and prompt adherence. We report mean opinion scores (MOS) on a 5-point scale. Audio quality and prompt adherence are rated from 1 (bad) to 5 (excellent), and diversity from 1 (identical) to 5 (diverse). All 14 test participants used a good playback system. 
{Given that in our objective evaluation we already assess broad sound synthesis capabilities, this qualitative evaluation is designed to explore specific use cases and challenges. We focus on prompts relevant to music production, such as ``latin funk drumset 115 BPM", as well as spatially complex scenes like ``sports car passing by". Additionally, to assess diversity, we include broader, more ambiguous prompts such as ``fire crackling" and ``water".}

\subsection{Baselines}
\label{sec:baseline}

\begin{itemize}
    \item \emph{Stable Audio Open (SAO)} \cite{evans2024open}: quality baseline and acceleration reference point, being both larger and not optimized for speed.
    \item \emph{Pre-trained RF} is our base accelerated model, see Sec. \ref{sec:rf}.
    \item \emph{Presto} \cite{Novack2025Presto} is a state-of-the-art {distillation} method for accelerating audio diffusion, using the base model and an auxiliary score model to minimize a reverse KL loss along with a GAN loss.
    \item \emph{Ablations.} %
    We ablate {ARC} by omitting $\mathcal{L}_C$ or replacing $\mathcal{L}_R$ with the standard least-squares adversarial ($\mathcal{L}_{\text{LS}}$) loss \cite{Novack2025Presto,postolache2023adversarial,guso2022loss}.
    \item We naively speed up methods by reducing the number of steps.
\end{itemize}

\noindent As, we aim to accelerate text-to-audio models without compromising quality. Hence, our qualitative evaluation includes the following baselines: Presto (8-step, distillation-based), SAO (100-steps, highest-quality), and pre-trained RF (50-step, before ARC post-training).

\subsection{Results and discussion}

Unsurprisingly, the best results are from the slow, much larger SAO~\cite{evans2024open}.
The studied accelerated models obtain comparable metrics but are 100x faster than SAO (100-steps), and 10x faster than the pre-trained RF model (50-steps). {We found significant differences in diversity MOS scores, but only minor, non-significant differences for quality/prompt adherence.}
One interesting outlier is Presto, which improves the quality of the base RF model (in terms of MOS score) at the expense of severely compromising diversity ({\color{red}red} in Table \ref{tab:main}) and worsening FD$_{\text{openl3}}$.
Conversely, ARC post-training further improves the diversity of the generations and, while obtains among the best FD$_{\text{openl3}}$, 
the MOS quality results are slightly worse than the pre-trained RF model. 
While Presto delivers high quality, its outputs lack variety, making it less suitable for generative tasks and creative applications. %
Our results also show that rectified flow models (pre-trained RF, 50-steps) are competitive with diffusion models (SAO, 100-steps) despite using half the steps. However, simply reducing the number of steps is not a viable solution, as it leads to significant quality degradation. 

Our first ablation shows that training with only $\mathcal{L}_R$ leads to bad prompt adherence. Interestingly, diversity scores are higher when prompt adherence is low ({\color{blue}blue} in Table \ref{tab:main}). This is because the generator becomes an unconditional model generating all types of (diverse) outputs, not following the prompt. 
In line with that, also note that ARC results show slightly lower prompt adherence (in MOS score) due to its improved diversity.
Our second ablation shows that the {relativistic} loss outperforms the least-squares loss for adversarial plus contrastive post-training. We also find that our model performs best using 8 steps, which is in line with recent work that smaller accelerated models may need more steps than their larger counterparts~\cite{geng2024consistency}.
Finally, we note that our proposed CCDS diversity metric is completely aligned with the results of the listening test, giving evidence that such a metric is reasonable in automatically assessing diversity.

\subsection{Edge-device optimizations}
\label{sec:arm}

We use Arm's KleidiAI library that is integrated into LiteRT runtime via the XNNPACK library. We experiment with a Vivo X200 pro phone with an Octa-core Arm CPU (1x Cortex-X925, 3x Cortex-X4, 4x Cortex-A720, 12GB RAM).
To balance quality and deployment efficiency, we chose dynamic Int8 quantization, which (when applied selectively) typically does not require quantization-aware training. Weights are quantized ahead of time, while activations are quantized dynamically at runtime based on their statistical distribution. This allows the model to benefit from reduced memory usage and faster inference while maintaining acceptable output quality. Since the quantization happens post-training and is applied only to certain layers, it offers a practical trade-off without significantly increasing development complexity.
This
decreases inference time from 15.3s (original F32) to 6.6s, and reducing peak runtime RAM usage from 6.5GB to 3.6GB. We can compare runtimes using high-end (H100) and consumer (3090) GPUs, which achieve speeds of 75ms and 187ms.

\subsection{Creative applications}

Our primary goal is to accelerate text-to-audio models for practical use in creative workflows. To feel like a compelling ``{instrument}", or have a similar experience, text-to-audio models must be responsive and react quick. To that end, we reduced latency to below 200ms on consumer-grade GPUs. We informally experimented with this model to make music and found it inspiring for sound design, due to its speed, prompt versatility, and capacity to generate unconventional sounds.
We were also captivated by its audio-to-audio capabilities for style transfer, which required no additional training. This is achieved by using any recording as the initial noisy sample $\mathbf{x}_{\tau_i}$ during ping-pong sampling (Sec. \ref{sec:sampling}). This approach enables voice-to-audio control by initializing $\mathbf{x}_{\tau_i}$ with a voice recording, as well as beat-aligned generations by initializing $\mathbf{x}_{\tau_i}$ with a recording having a strong beat. We provide examples on our demo page. %
However, a key limitation of our model is its memory and storage requirements, occupying several GB of RAM and disk space (Sec. \ref{sec:arm}), which may pose challenges for integration into many applications and for its efficient distribution.

\section{Conclusion}

ARC post-training is the first acceleration method for text-to-audio models that does not rely on distillation or CFG. By extending an adversarial relativistic loss to gaussian flow acceleration, combined with a novel contrastive discriminator loss, we speed up gaussian flow models' 
runtime to milliseconds, with consumer-grade GPUs, or to seconds, with edge-device CPUs. Such speedups are obtained without significantly compromising quality and increasing the diversity of the generations. Diversity is measured using our proposed CCDS metric, which we find to align with perceptual assessments of diversity.
We hope that, with improved efficiency and diversity, text-to-audio models will soon be able to support a broader range of creative applications. Recognizing the creative potential of such models, we also include a small audio-to-audio experimentation, and future efforts may focus on fine-tuning with targeted datasets for more precise sound design.

\bibliographystyle{IEEEtran}
\bibliography{refs25}

\end{document}